\documentclass[apj,twocolumn]{myemulateapj}
\pdfoutput=1
\journalinfo{The Astrophysical Journal, \href{http://iopscience.iop.org/0004-637X/753/1/4/}{753:4 (9pp), 2012 July 1}}
\doiinfo{doi:~\href{http://dx.doi.org/10.1088/0004-637X/753/1/4}{10.1088/0004-637X/753/1/4}}
\submitted{Received 2012 February 22; accepted 2012 April 23; published 2012 June 7}
\copyrightinfo{}
\usepackage{graphicx}
\usepackage[pdftitle={Quantifying the Biases of Spectroscopically Selected Gravitational Lenses},pdfauthor={Ryan A. Arneson, Joel R. Brownstein, Adam S. Bolton}, bookmarks, bookmarksopen, colorlinks=true, linkcolor=blue, citecolor=blue, anchorcolor=blue, urlcolor=blue]{hyperref}

\shorttitle{Quantifying the Biases of Spectroscopically Selected Gravitational Lenses}
\shortauthors{Arneson, Brownstein, \& Bolton}
\newcommand{\sref}[1]{Section \ref{#1}}
\begin{document}
 
\title{Quantifying the Biases of Spectroscopically Selected Gravitational Lenses}

\author{\mbox{Ryan A. Arneson\altaffilmark{1,2}}, \mbox{Joel R. Brownstein\altaffilmark{1}}, and \mbox{Adam S. Bolton\altaffilmark{1}}}
\affiliation{\(^1\)Department of Physics and Astronomy, University of Utah, Salt Lake City, UT 84112, USA.}
\affiliation{\(^2\)Department of Physics and Astronomy, University of California at Irvine, Irvine, CA 92697, USA.}
\email[]{arnesonr@uci.edu}
\email[]{joelbrownstein@astro.utah.edu}
\email[]{bolton@astro.utah.edu}

\begin{abstract}
Spectroscopic selection has been the most productive technique for the selection of galaxy-scale strong gravitational lens systems with known redshifts.  Statistically significant samples of strong lenses provide a powerful method for measuring the mass-density parameters of the lensing population, but results can only be generalized to the parent population if the lensing selection biases are sufficiently understood.  We perform controlled Monte Carlo simulations of spectroscopic lens surveys in order to quantify the bias of lenses relative to parent galaxies in velocity dispersion, mass axis ratio, and mass-density profile.  For parameters typical of the SLACS and BELLS surveys, we find (1) no significant mass axis ratio detection bias of lenses relative to parent galaxies; (2) a very small detection bias toward shallow mass-density profiles, which is likely negligible compared to other sources of uncertainty in this parameter; (3) a detection bias towards smaller Einstein radius for systems drawn from parent populations with group- and cluster-scale lensing masses; and (4) a lens-modeling bias towards larger velocity dispersions for systems drawn from parent samples with sub-arcsecond mean Einstein radii.  This last finding indicates that the incorporation of velocity-dispersion upper limits of \textit{non-lenses} is an important ingredient for unbiased analyses of spectroscopically selected lens samples.  In general we find that the completeness of spectroscopic lens surveys in the plane of Einstein radius and mass-density profile power-law index is quite uniform, up to a sharp drop in the region of large Einstein radius and steep mass-density profile, and hence that such surveys are ideally suited to the study of massive field galaxies.
\end{abstract}

\keywords{galaxies: statistics -- gravitational lensing: strong -- surveys}

\slugcomment{To be submitted to The Astrophysical Journal}

\maketitle

\section{Introduction}
\label{Intro}
Galaxy-scale strong gravitational lenses represent a uniquely precise tool for measuring the mass-density parameters of galaxies at cosmological distances, and their importance to the study
of galaxy structure and evolution continues to increase as known examples become more numerous.
Within this area, the method of spectroscopic gravitational lens selection stands out for its productivity and for its ability to deliver precise redshifts of
both components in a lens system.  In this method, lens systems are found through data mining procedures within large spectroscopic galaxy redshift survey databases that identify at least one emission-line (such as [O\,{\sc ii}]\,$\lambda$3727) in a target galaxy's spectrum that originates from a more distant background source.
In the context of spectroscopic lens surveys such as the Sloan Lens ACS Survey \citep[SLACS;][]{Bolton:2006ApJ...638..703B,bolton08,Auger:2009ApJ...705.1099A}, the Optimal Line-of-sight Survey \citep[OLS;][]{willis05,willis06}, the Sloan WFC Edge-on Late-type Lens Survey \citep[SWELLS;][]{swells1}, the BOSS Emission-Line Lens Survey \citep[BELLS;][]{bells1}, and possible similar studies within future projects such as BigBOSS~\citep{Schlegel.2009arXiv0904.0468S}, it has become imperative to understand the spectroscopic lens selection function, in order to draw accurate conclusions about the mass-density parameters of the parent population.

\citet{Kochanek.1992ApJ...397..381K} and \citet{Mortlock.2000MNRAS.319..879M} have made predictions for the frequency of lensed quasars discoveries in galaxy redshift surveys, and \citet{Mortlock.2001MNRAS.321..629M} emphasized the critical role of the finite radius of the spectroscopic fiber in the selection of lensed quasars.  A detailed investigation of the selection function of spectroscopically discovered
gravitational lensed galaxies was made by \citet{dobler08}.  That work focused
primarily on predicting the overall incidence of lenses within the Sloan Digital
Sky Survey \citep[SDSS;][]{york00} spectroscopic database from which the
SLACS lenses were selected. 

Our current work is distinguished by the use of
a more accurate [O\,{\sc ii}] luminosity function, a focus
on the \textit{relative} incidence of lenses as a function of other physical parameters
and the associated selection and modeling biases,
and a consideration of the effects of variations in the radial gradient of
the mass-density profile of the lens galaxies.
Our goal is to quantify the parameter distribution biases of spectroscopic lens samples
relative to the parent galaxy samples from which they are selected.
In contrast to \citet{dobler08}, we rely almost exclusively on
Monte Carlo methods as opposed to analytic methods.
Our current work is also complementary to the work of \citet{vandeven09}
and \citet{mandelbaum09}, which quantify the (much different) selection
biases of lenses discovered within imaging survey data.

Our approach in this paper is to use controlled Monte Carlo simulations to generate mock lens and source catalogs with realistic parameters, to numerically analyze the magnification of these systems using ray-tracing techniques, then to apply SDSS-like spectroscopic selection cuts as well as modeling constraints to determine which mock lens and source parameters are preferentially selected and modelable as strong gravitational lens systems.  Our aim is to allow the selection biases in spectroscopic galaxy-scale strong lens surveys to be understood and corrected, and to aid in the design of future surveys and follow-up observations.

This paper is structured as follows.  We describe the Monte Carlo simulation method and generation of the mock catalogs in \sref{simulation}.  In \sref{results}, we present the results of our simulations.  We discuss these results and draw conclusions in \sref{conclusion}.
We assume a general-relativistic Friedmann--Robertson--Walker  (FRW) cosmology with matter-density parameter  $\Omega_{m}\!=\!0.3$, vacuum energy-density parameter $\Omega_{\Lambda}\!=\!0.7$, and Hubble parameter $H_o\,=\,70\ {\rm km\,s}^{-1}\,{\rm Mpc}^{-1}$ throughout this paper.
(Note however that our results are almost completely insensitive to the detailed values of these cosmological
parameters.)

\section{Monte Carlo Simulation}
\label{simulation}

In this section we describe our Monte Carlo lens system simulation procedure in detail.
An overview of the steps is as follows:
\begin{enumerate}
\item generate a mock catalog of lens galaxy parameters,
\item generate a mock catalog of source galaxy parameters,
\item calculate the magnification of the source galaxy observed within an SDSS fiber,
\item randomly select an [O\,{\sc ii}] luminosity for the source galaxy,
\item calculate the [O\,{\sc ii}] line flux in the fiber,
\item determine if the source line flux is detected, and
\item determine if the lens system is modelable.
\end{enumerate}
These steps were repeated over a sequence of runs for two representative background source redshifts $z_{s}$ and three representative mean Einstein radius values $\overline{\theta_E}$
described in further detail below.  The first three steps define a particular lens geometry,
and each run includes 1 million random realizations.  Each of these realizations was associated with
1 million random draws from the luminosity function, leading to $10^{12}$ effective realizations
per run.  The simulations were carried out over several days of run time on a computer cluster
with 8 nodes of 12 processor cores each.

\subsection{Lens Galaxy Catalog}
The singular power-law ellipsoid~\citep[SPLE;][]{Barkana.1998ApJ...502..531B} was chosen as the parametric model for the mock lens galaxy catalog.  The SPLE is a generalization of the singular isothermal ellipsoid~\citep[SIE;][]{Kassiola.1993ApJ...417..450K,Kormann.1994AA...284..285K,Keeton.1998ApJ...495..157K} by means of an additional parameter, $\gamma$, which describes the mass-density power-law index, where $\rho \propto r^{-\gamma}$.   A power-law index of $\gamma = 2.0$ corresponds to a SIE model in the circular limit.  A $\gamma$ greater than (or less than) two corresponds to a steeper (or shallower) lensing potential.

The parameters we consider in the creation of our mock lens galaxy catalog included: the mass normalization of the lensing galaxy (i.e. the Einstein radius, $\theta_E$), the power-law index $\gamma$ of the mass-density profile, and the projected mass axis ratio $q$.  It is important to point out that although in this paper we use Einstein radius as a proxy for the mass of the lens galaxy it does not imply only circularly symmetric lens configurations or Einstein rings.  In the limit of a singular isothermal sphere (SIS), the Einstein radius is related to the physical mass through
\begin{equation}
\label{radius}
\theta_E = 4\pi\frac{{\sigma_{v}}^2}{c^2}\frac{D_{LS}}{D_S},
\end{equation}
where $\sigma_v$ is the velocity dispersion of the lens galaxy, $D_{LS}$ and $D_S$ are cosmological angular-diameter distances from lens to observer and observer to source, respectively.  Based on this relationship, $\theta_E$ can be considered as a stand-in for $\sigma_{v}$ at fixed lens and source redshift.  Following \citet{Shu.2012AJ....143...90S} who find a typical rms dispersion of 0.075 dex
in $\sigma_{v}$ at fixed luminosity
and redshift for massive galaxies at redshift $z \sim 0.5$, we adopt a log-normal Einstein radius distribution in our simulations with an intrinsic rms dispersion of 0.15 dex.  Since we randomly draw from a distribution in $\theta_E$ rather than $\sigma_v$, the simulation itself is independent of lens galaxy redshift.  We explore the parameter space by considering log-normal distributions centered on Einstein radii of $\overline{\theta_E} = 0\farcs5$, $1\farcs0$, and $2\farcs0$ in separate simulations, thus spanning the representative range of galaxy-scale strong lensing systems.
The mass-density power-law indices of the SPLE lens galaxies are drawn from a Gaussian distribution, with $\gamma = 2.0\pm0.2$, further constrained to be between $1.0 \le \gamma \le 3.0$, throughout.  We adopt this distribution as it is both representative of the SLACS lens galaxies~\citep{koopmans06} and capable of sampling parameter space.
The mass axis ratios are drawn from a distribution which approximately replicates the results found in \citet[Figure 6]{holden09} for the ellipticities of early-type galaxies from $z\sim0$ to $z\sim1$.  The simulated lens galaxies are placed at the center of the simulated spectroscopic fiber aperture, and an intermediate axis convention is used for Einstein radius normalization of elliptical lens models.  There is no need to randomly orient the position angle of the lenses, as the simulated source galaxies are randomly positioned.

\subsection{Source Galaxy Catalog}
In the creation of the mock source galaxy catalog we consider: the source scale radius, $r_s$, the position in the fiber, and the source redshift, $z_s$.  The surface density profile of the source is assumed to be an exponential disk and the scale radii of the sources are assumed to be uniformly distributed between $0\farcs005 \le r_s \le 1\farcs0$.  The lower limit is chosen to correspond to a source with a radius of half a pixel in the pixelized source plane of the simulation.
Source positions are uniformly distributed within a circular region of radius
2.5 times the spectroscopic fiber radius centered
on the lens in the source plane.
A radius significantly larger than the fiber is necessary to obtain
sufficient representation of unlensed sources at large radii that
nevertheless contribute detectable line flux within the fiber.
We represent SLACS-like systems by setting the source redshift to be $z_s = 0.6$,
based on the source redshift distribution of $0.63\pm0.20$ found for the SLACS grade-A lenses. 
As our simulations are designed, the value of $z_s$ only affects the detailed
form of the source luminosity function (below) and the projection of the observational
detection threshold within it.

\begin{figure*}[t]
\begin{center}
\includegraphics[scale=.95]{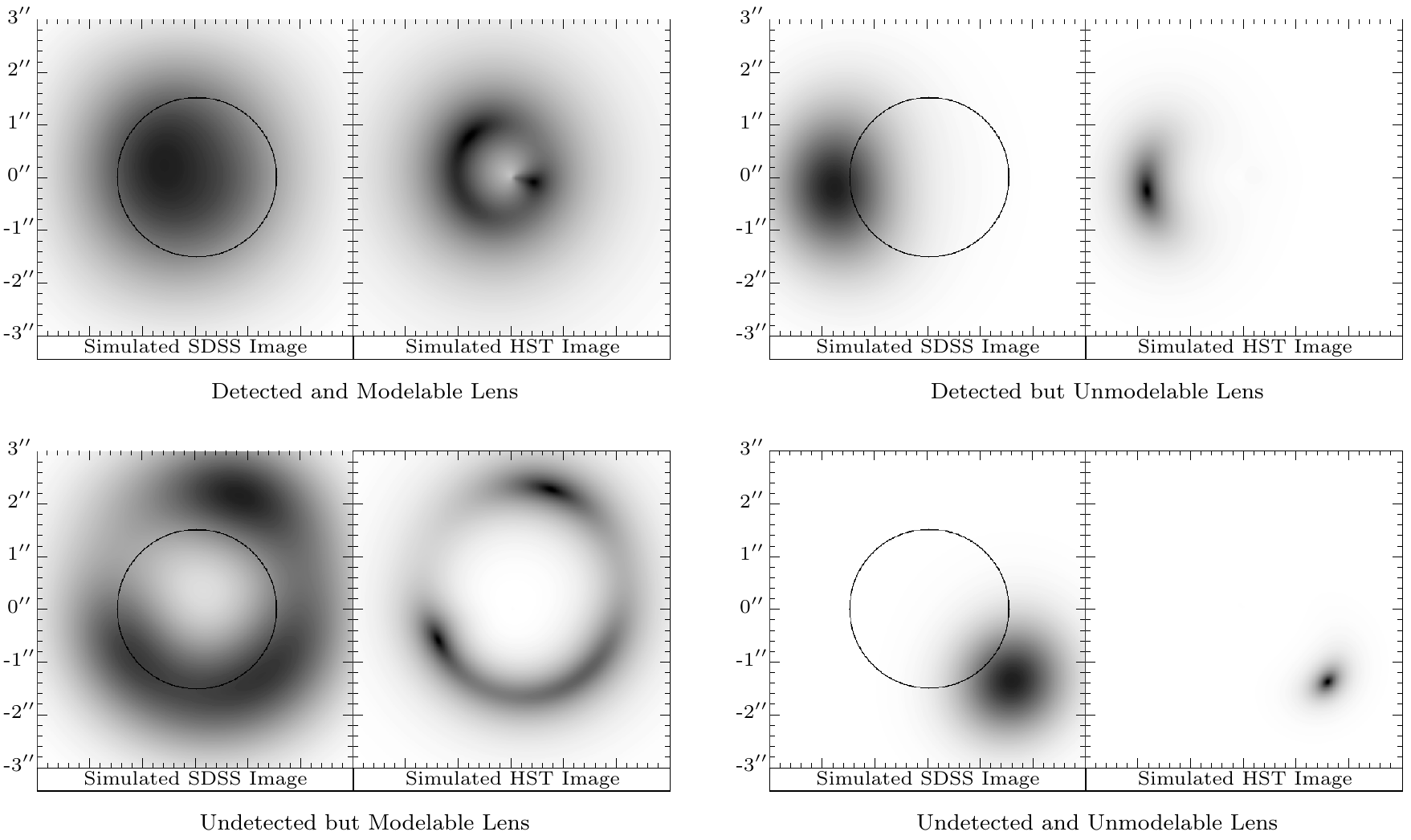}
\end{center}
\figcaption{\label{sim_ims}Example lens images from the mock SLACS simulations as would be seen by the SDSS (left) and {\em Hubble Space Telescope} ({\em HST}; right).  The circle represents an SDSS fiber ($3\farcs0$ in diameter).  The seeing was taken to be 1\farcs5 (FWHM) for the simulated SDSS images and 0\farcs05 (FWHM) for the simulated {\em HST} images.}
\end{figure*}

\subsection{System Catalog}
In order to create simulated lens system, each of the lens galaxy parameters from the lens catalog were pair-wise randomly associated with each of the parameters from the source galaxy catalog. 
This leads to a variety of lensed image configurations: doubles, three-image naked cusps, quads, rings, and of course a large number of singly-imaged sources.

\subsection{Source Magnification}
\label{magnification}
Here we summarize the method used to calculate the magnification of the source line flux as seen through the spectroscopic fiber.  In lens surveys such as SLACS (or BELLS), it is critical to account for the finite solid-angle of the original SDSS fiber of radius $1\farcs5$, or the upgraded SDSS-III BOSS fiber of radius $1\farcs0$~\citep{Eisenstein:2011AJ....142...72E}.  The strength of the source emission-line is directly related to how much line flux from the source galaxy is received into the fiber before it is sent to the spectrographs.  If the lens galaxy has an Einstein radius larger than the radius of the fiber it is possible that no, or very little, source flux will be received by the fiber and the lens system will fail to be detected.  This doesn't mean, however, that large Einstein radius lenses can't be spectroscopically detected.  Seeing effects also need to be accounted for, especially because the seeing full-width at half-maximum (FWHM) is as large (or larger than) the fiber radius.  Seeing effects can either add flux to the fiber if the source image is outside the fiber or remove flux from the fiber if the source image is inside the fiber, as demonstrated in Figure~\ref{sim_ims}. 

The fiber magnification is defined to be
\begin{equation}
\label{mu}
\mu_{\rm fib}=\frac{f_r}{f_i},
\end{equation}
where $f_r$ is the lensed source flux received by the fiber and $f_i$ is the intrinsic (unlensed) source flux that would have been received in the absence of lensing and fiber effects.  The lensed source flux $f_r$ (in arbitrary units) received by the fiber is calculated by first generating a $8\arcsec \times 8\arcsec$ ($800^2$ pixels) lensed image by ray-tracing the source plane image through the analytic lens mass model and subsequently integrating over a fiber mask convolved with the SDSS seeing PSF ($1\farcs5$ FWHM) centered on the lens galaxy.  This second step is mathematically equivalent, and computationally faster, to convolving the lensed image with the seeing PSF and then applying the top-hat fiber mask.  The intrinsic source flux $f_i$ (in the same arbitrary units)
is found by computing and integrating an unlensed image of the same parameterized source
galaxy, and the scalar magnification $\mu_{\rm fib}$ is given by the resulting ratio.
The computation of simulated lensed images is the most computationally intensive
aspect of our procedure.

\begin{figure*}[t]
\plottwo{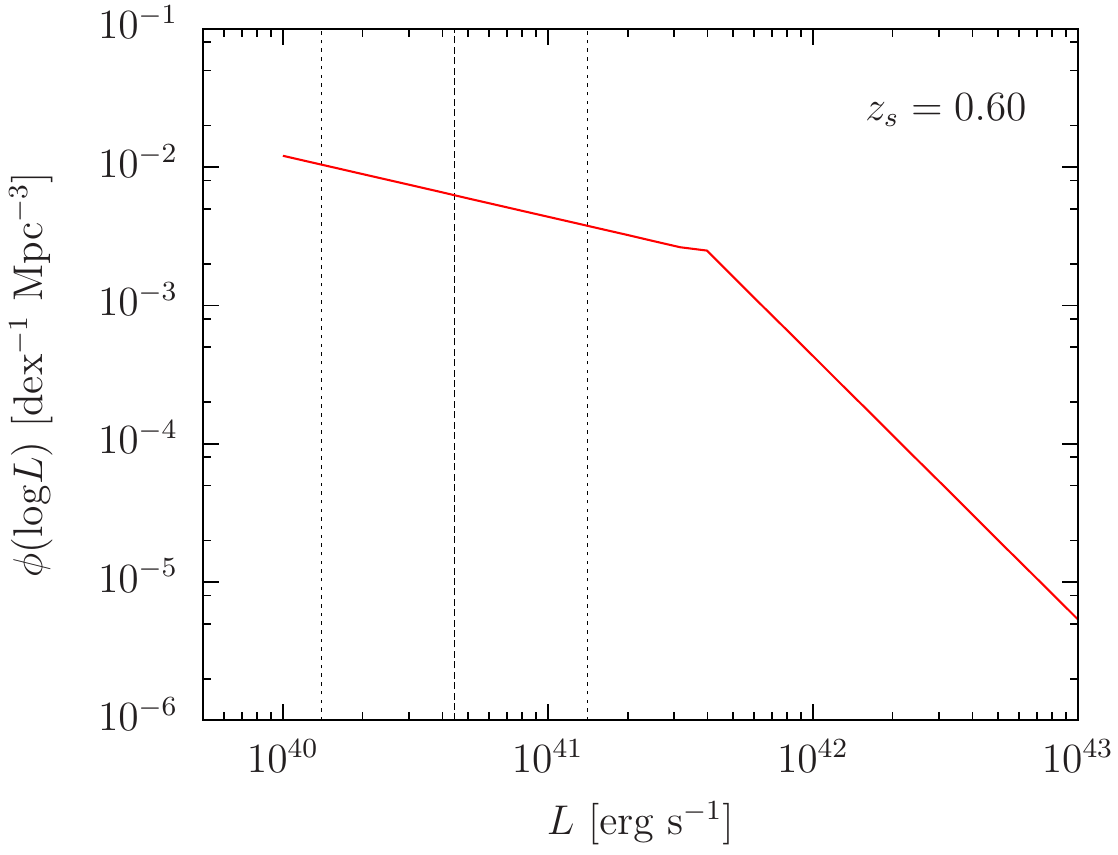}{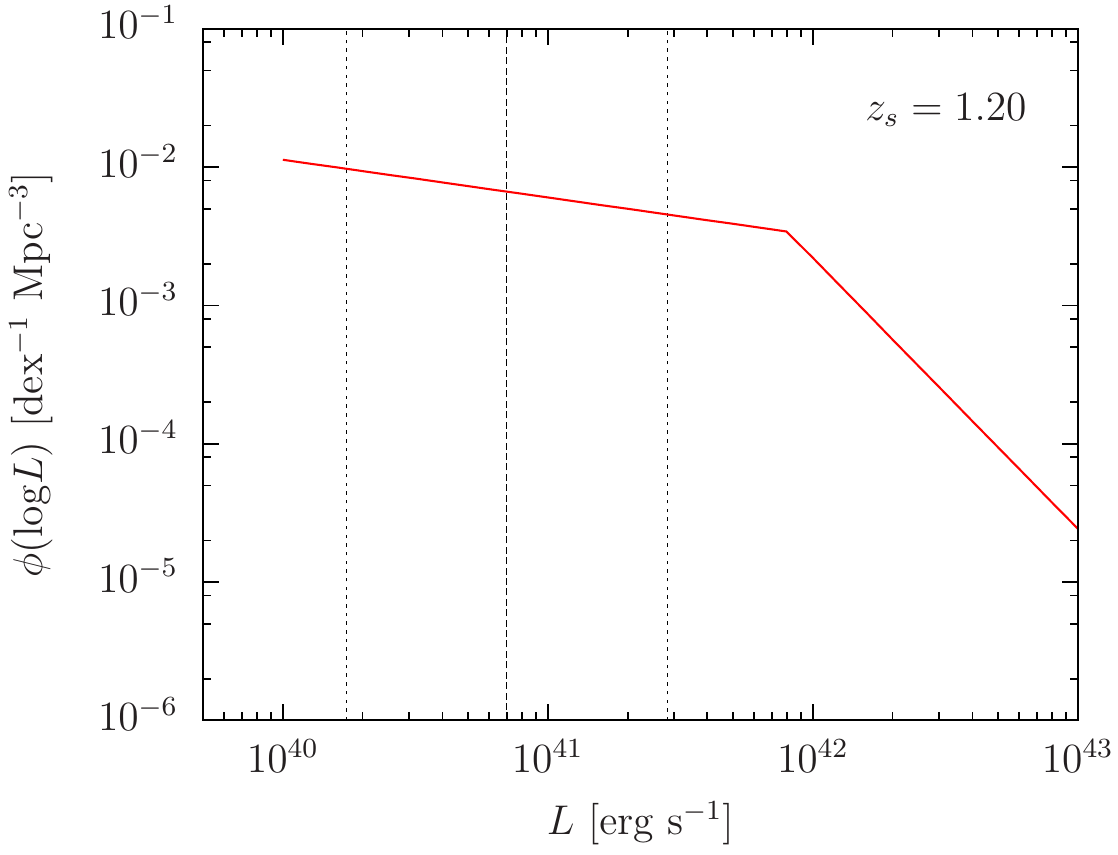}
\figcaption{\label{LF} [O\,{\sc ii}] luminosity functions (red) from which the source luminosities were drawn in the simulation.  The left and right panels represent the luminosity function at $z_s=0.60$ and $z_s=1.2$, respectively.  The lowest luminosity considered was $10^{40}\ {\rm erg\,s}^{-1}$.  The dashed lines represent the calculated average source luminosity for each redshift, the dotted lines are the calculated $1\sigma$ deviations in those luminosities.  The luminosity function at ${z_s=0.60}$, $(\alpha_f,\alpha_b,\log\,L_{\rm{TO}})=(-1.46,-2.90,41.50)$ and at ${z_s=1.2}$, $(\alpha_f,\alpha_b,\log\,L_{\rm{TO}})=(-1.27,-2.96,41.91)$.}
\end{figure*}

\subsection{[O\,{\sc ii}] Luminosity and Line Flux}\label{O2}

The [O\,{\sc ii}] luminosity function for our simulations is interpolated between low redshifts, using the results of \citet{gilbank10}, and redshifts $z\sim1$, using the results of  \citet{zhu09}.  In both papers, the [O\,{\sc ii}] luminosity function is fitted by a double power law:
\begin{equation}
\label{phi}
\phi(L_{\rm{O}_{\rm{II}}},z_s) = \frac{dN}{dL_{\rm{O}_{\rm{II}}}} \propto {L_{\rm{O}_{\rm{II}}}}^\alpha,
\end{equation}
where $L_{\rm{O}_{\rm{II}}}$ is the [O\,{\sc ii}] luminosity and $\alpha$ is a dimensionless parameter.  Thus, we interpolate the turnover luminosity ($L_{\rm{TO}}$) in the power-law function as well as the slopes at brighter ($\alpha_{b}$) and fainter ($\alpha_f$) luminosities from $L_{\rm{TO}}$ as a function of redshift.  Each source galaxy in a particular lens configuration is assigned $10^{6}$ randomly drawn [O\,{\sc ii}] luminosities according to the interpolated luminosity function, which in turn depends on the assumed source redshift.  The [O\,{\sc ii}] luminosity functions for both source redshifts considered in this work are presented in Figure~\ref{LF}. Note that the source size and [O\,{\sc ii}] luminosity are assumed to be uncorrelated.

\begin{figure}[t]
\begin{center}
\includegraphics[scale=.825]{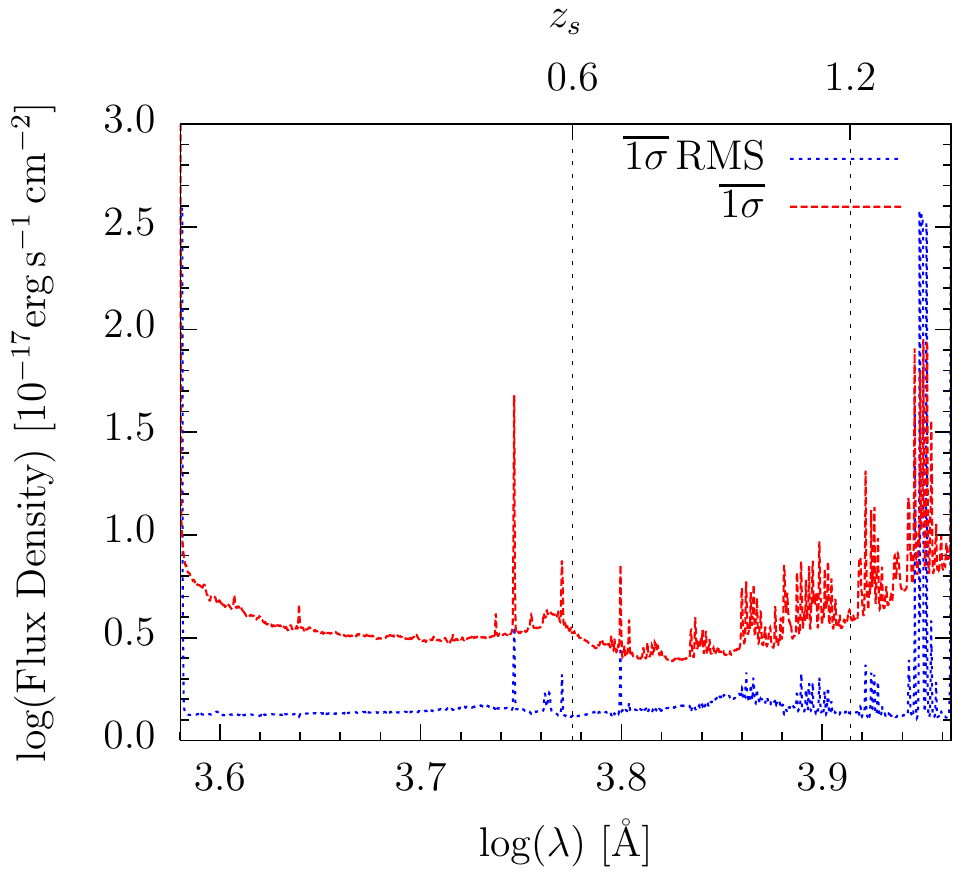}
\end{center}
\figcaption{\label{residual} Log median $1\sigma$ [O\,{\sc ii}] line flux, $\overline{1\sigma}$, detection limit spectrum for SDSS (dashed red) and the log [O\,{\sc ii}] $\overline{1\sigma}$ rms (dotted blue), each as a function of observed log [O\,{\sc ii}] wavelength, on the lower horizontal scale, and simulation redshifts, on the upper horizontal scale.  The values were compiled from $\sim~8 \times 10^5$ spectra from the SDSS DR7.  The dashed vertical lines show where the sources tested by the simulation lie within the noise spectrum.}
\end{figure}
\vspace{3.0mm}

The simulated [O\,{\sc ii}] line flux for each realization,
$S_{\rm{O}_{\rm{II}}}$, is given by the equation:
\begin{equation}
\label{s}
S_{\rm{O}_{\rm{II}}} = \frac{L_{\rm{O}_{\rm{II}}}\,\mu_{\rm fib}}{4\,\pi\,{D_L}^2},
\end{equation}
where $L_{\rm{O}_{\rm{II}}}$ is the [O\,{\sc ii}] luminosity, $\mu_{\rm fib}$ is the fiber magnification, and $D_L$ is the cosmological luminosity distance from observer to source.  After calculating the line flux received by a fiber, the detection classification of a realization is dependent on the delivered line flux relative to the noise limit of the SDSS spectroscopy.  In order to determine this detection limit, the median and rms $1\sigma$ line flux for the SDSS spectroscopic database is calculated from $\sim 8\times 10^5$ spectra from the SDSS seventh data release ~\citep[DR7;][]{SDSS:2009ApJS..182..543A}.  Figure~\ref{residual} shows the median $1\sigma$ line flux and rms for SDSS.  A $1\sigma$ line-flux noise limit for the [O\,{\sc ii}]  emission-line at the source redshift is drawn from this normal distribution for each realization.  Based on the $6\sigma$ [O\,{\sc ii}]  flux detection limit applied to the discovery of SLACS and BELLS lenses,  the simulated source is considered to be detected if the calculated [O\,{\sc ii}]  line flux of the catalog source is greater than six times the randomly drawn $1\sigma$ value.

\subsection{Modelable Constraints}

The criteria used for determining the modelability of each lens system were estimated from the results of SLACS and BELLS grade-A lens models.
We consider a system to be ``modelable'' as a strong lens if its Einstein radius is greater than the minimum Einstein radius of the lenses modeled by \citet{bells1} and if
the source position is within the radial caustic of the SIE lens model.
Numerically, these requirements are that $\theta_E > 0\farcs5$ and that the source
position is less than $2\theta_E$ from the center of the lens galaxy.  This second limit corresponds to the point at which a counter image of the source appears in the image plane in the case of an SIS, and encodes a requirement for multiple imaging in order to classify a system as a lens and to fit a strong-lens model.   Although not all the lenses are SISs, this constraint is reasonably general.
Note that a lens configuration is classified as either modelable or not depending upon its geometry,
but independent of its [O\,{\sc ii}] line flux.
For a given lens configuration, the multiple [O\,{\sc ii}] luminosity realizations
allow us to compute a fractional detectability value.
Lenses that are detected but not modelable correspond to ``false positive'' candidates,
which nevertheless play an important role in the population distribution
considerations to be discussed further below.

\subsection{Mock BELLS Simulation}
\begin{deluxetable*}{cc|cc|ccccc|ccccc}
    \tabletypesize{\scriptsize}
    \tablewidth{\hsize}
    \centering
    \tablecaption{\label{table}Mock Simulation Parameters with Detected and Modelable Results}
    \tablehead{
      \multicolumn{2}{c|}{} & \multicolumn{2}{c|}{Parent} & \multicolumn{5}{c|}{Detected} & \multicolumn{5}{c}{Modelable}\\ \tableline
      Run &
      {$z_s$} & 
      {${\overline{\theta_E}}$} &
      {$\sigma (\theta_E)$} &
      {$\overline{\theta_E}$} & 
      {$\sigma (\theta_E)$} &
      $\phantom{\bigr(}$ $\gamma$ $\phantom{\bigr)}$ &
      Number &
      Percent &
      {$\overline{\theta_E}$} & 
      {$\sigma (\theta_E)$} &
      $\gamma$ &
      Number &
      Percent
      \\
      &
      &
      [$\arcsec$] & 
      [dex] &
      [$\arcsec$] & 
      [dex] &
      &
      &
      &
      [$\arcsec$] & 
      [dex] &
      &
      &
}
    \tablecolumns{14}
    \startdata
        \(1\) & \(0.60\) & \(0.5\) & \(0.15\) & \(0.51\) & \(0.15\) & \(1.99 \pm 0.20\) & \(8.42 \times 10^{3}\) & \(0.84\%\) & \(0.64\) & \(0.09\) & \(1.99 \pm 0.20\) & \(3.59 \times 10^{3}\) & \(42.61\%\) \\ 
        \(2\) & \(0.60\) & \(1.0\) & \(0.15\) & \(1.01\) & \(0.15\) & \(1.99 \pm 0.20\) & \(9.06 \times 10^{3}\) & \(0.91\%\) & \(1.03\) & \(0.14\) & \(1.99 \pm 0.20\) & \(8.39 \times 10^{3}\) & \(92.67\%\) \\ 
        \(3\) & \(0.60\) & \(2.0\) & \(0.15\) & \(1.87\) & \(0.14\) & \(1.97 \pm 0.20\) & \(7.74 \times 10^{3}\) & \(0.77\%\) & \(1.87\) & \(0.14\) & \(1.97 \pm 0.20\) & \(7.70 \times 10^{3}\) & \(99.54\%\) \\ \tableline
        \(4\) & \(1.20\) & \(0.5\) & \(0.15\) & \(0.52\) & \(0.15\) & \(1.98 \pm 0.20\) & \(1.85 \times 10^{3}\) & \(0.19\%\) & \(0.64\) & \(0.09\) & \(1.98 \pm 0.20\) & \(9.60 \times 10^{2}\) & \(51.86\%\) \\ 
        \(5\) & \(1.20\) & \(1.0\) & \(0.15\) & \(0.99\) & \(0.14\) & \(1.97 \pm 0.20\) & \(2.05 \times 10^{3}\) & \(0.20\%\) & \(1.00\) & \(0.13\) & \(1.97 \pm 0.20\) & \(1.97 \times 10^{3}\) & \(96.42\%\) \\ 
        \(6\) & \(1.20\) & \(2.0\) & \(0.15\) & \(1.71\) & \(0.13\) & \(1.93 \pm 0.20\) & \(1.24 \times 10^{3}\) & \(0.12\%\) & \(1.71\) & \(0.13\) & \(1.93 \pm 0.20\) & \(1.24 \times 10^{3}\) & \(99.89\%\)  
    \enddata
    \tablecomments{The Run number reflects the results found in Figure~\ref{histogram} with Runs 1--3 and Runs 4--6 corresponding to the mock SLACS and BELLS simulations, respectively.  $z_s$ represents the source redshift.  The mean Einstein radii are given by $\overline{\theta_E}$ (and in fact represent exponentiated mean log Einstein radii).  The standard deviation of the Einstein radii, $\sigma (\theta_E)$, are in units of dex.  The values of $\overline{\theta_E}$ and $\sigma (\theta_E)$ presented for the Parent sample are the input values upon which the distributions are drawn.  All of the Parent samples had mass density power-law index, $\gamma$, distributions with a mean of 2.0 and a standard deviation of 0.2.  The values of $\overline{\theta_E}$, $\sigma (\theta_E)$, and $\gamma$ for the detected and modelable lenses were calculated by fitting a gaussian function to the binned histograms.  Each run included \(10^6\) mock lens systems in the Parent samples.  The ``Number'' columns indicate the number of mock lens systems generated by the Monte Carlo simulation which were detected, and the ``Number'' determined to be modelable.  The ``Percent'' of detected lenses is relative to the Parent sample of \(10^6\) systems, whereas the ``Percent'' of modelable lenses is relative to the Detected sample.}
\end{deluxetable*}

In order to extend the simulation and to understand the effects that survey-specific parameters have on the selection biases, we also run the simulation with parameters typical of the BELLS survey.  We decrease the fiber radius to $1\farcs0$ to conform to the SDSS-III BOSS fiber upgrade~\citep{Eisenstein:2011AJ....142...72E}.   To replicate the depth of the BELLS survey, from which 25 definite and 11 probable strong galaxy--galaxy lenses have been discovered with source redshifts, $z_s = 1.2\pm0.2$~\citep{bells1}, we place the sources at redshift of $z_s=1.2$.  Although a rigorous treatment of the mock BELLS simulation would recalculate the median $1\sigma$ [O\,{\sc ii}] line flux of Figure~\ref{residual} specifically for BOSS spectra, we leave this for future work; for our purposes here, we simply lower the detection criteria from $6\sigma$ to $4\sigma$ to mimic the anticipated fainter detection threshold of BOSS spectroscopy.  Again we explore parameter space by considering the same mean Einstein radii of $\overline{\theta_E} = 0\farcs5$, $1\farcs0$, and $2\farcs0$ for the parent distributions; all other parameter distributions were left unchanged.

\subsection{Completeness Simulation}

In addition to the simulations above, which aim to quantify parameter biases
in lens samples selected from realistic lens and source distributions,
we carry out two simulations with uniform input distributions in the
Einstein radius $\theta_E$ and mass-density slope parameter $\gamma$,
in order to assess spectroscopic selection completeness in the
plane spanned by these two parameters.  We compute $2 \times 10^6$
realization for both SLACS-like and BELLS-like parameters,
and quantify both the detected and detected$+$modelable
fractions as a function of these parameters.

\section{Results}
\label{results}

\begin{figure*}[t]
\begin{center}
\includegraphics[scale=.635]{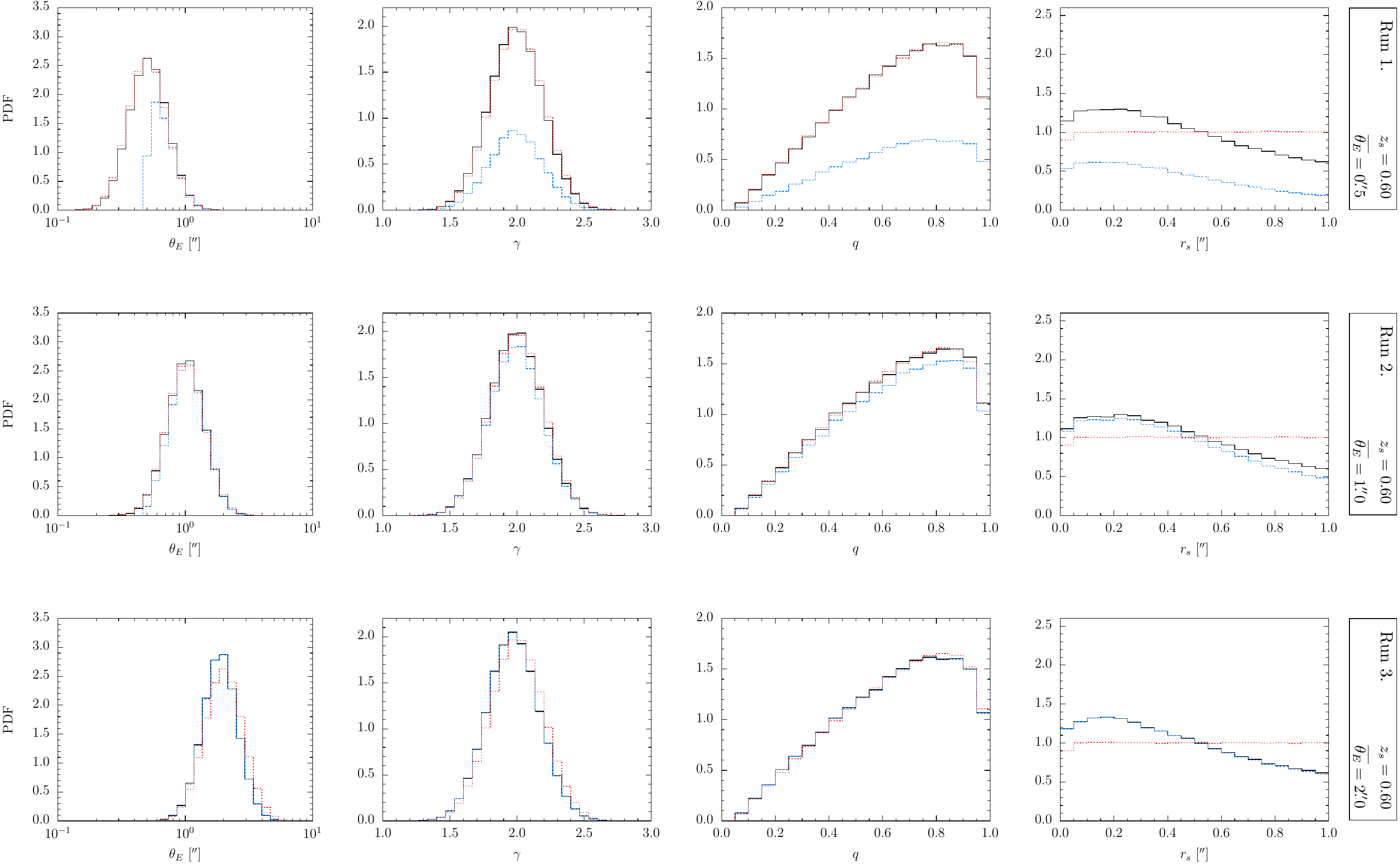}
\end{center}
\vspace{-5mm}
\begin{center}
\includegraphics[scale=0.635]{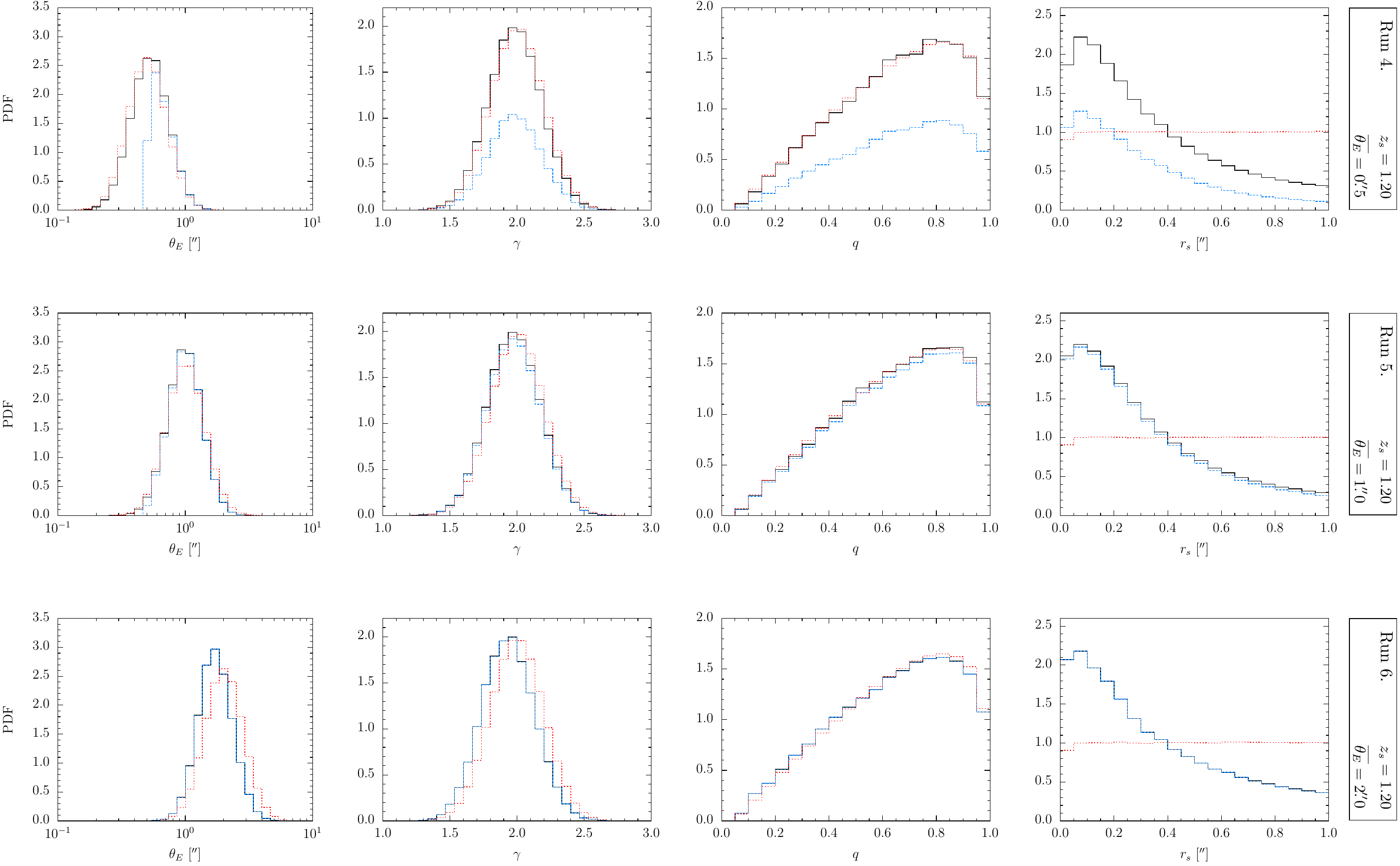}
\figcaption{\label{histogram}Einstein radius ($\theta_E$), lens mass-density power-law index ($\gamma$), lens mass axis ratio (q), and effective source scale radius ($r_s$) of the parent (dashed red line), detected (black line), and modelable (dashed blue line) lens samples for each run of the simulation, as summarized in Table \ref{table}, arranged by source redshift.  From the top to bottom row, we vary the parent mean Einstein radius, $\overline{\theta_E}$, between $0\farcs5$, $1\farcs0$, and $2\farcs0$, respectively.  The parent sample consisted of \(1 \times 10^6\) mock lenses.  The parent and detected histograms are normalized to unit area, whereas the modelable histogram is normalized to the fraction of modelable lenses relative to detected lenses.}
\end{center}
\end{figure*}

In this section we present the results of the Monte Carlo simulation, outlined in \sref{simulation}, used to generate \(1\times 10^6\) lens systems per run.  The results for each run are summarized in Table~\ref{table}.  Figure~\ref{histogram} depicts the histograms of the Einstein radius, lens mass-density power-law index, lens mass axis ratio, and effective source size for the parent, detected, and modelable lenses for each of the runs.  Note that by ``modelable'' we are referring to the modelable subset of \textit{detected} systems (since this is the experimentally relevant definition), even though it is possible for an undetected system to be modelable in its geometry.  As suggested by the procedure discussed in \sref{simulation}, the lens selection biases can be divided into a detection bias and a modeling bias.  Throughout this section both biases will be presented side-by-side with respect to a given parameter.  We first discuss the results of the SLACS-like simulations.

\subsection{Dependence on Lens Parameters}
The mean Einstein radius (or rather, the exponentiated mean log Einstein radius) and standard deviation for the detected and modelable samples were calculated by fitting a Gaussian function to the binned histograms displayed in Figure~\ref{histogram}.
The only appreciable bias in detected $\theta_E$ shows up for the simulated
population with a mean $\overline{\theta_E}= 2\farcs0$, due to the finite size
of the fiber.  As the Einstein radius or strength of the lens increases the chance that enough of the lensed image falls inside the fiber to be detected decreases. And, as mentioned in \sref{magnification}, even though the seeing effects may introduce more flux into the fiber for images outside the fiber it simply may not be enough for the system to be detected.  The only hope for a large Einstein radius lens to be detected is if it has a bright counter image that lies within the fiber, which is possible but not probable.  The fiber still imparts a bias against large separation lenses.
The bias in Einstein radius for the modelable lenses, by contrast, is most appreciable for parent samples with the small mean Einstein radius of $\overline{\theta_E}= 0\farcs5$.
This results from the requirement that $\theta_E \ge 0\farcs5$, which eliminates small radius lenses and increases the average modelable Einstein radius.

The overall percentage of detected systems is small --- less than 1\% --- and is
maximized for the parent sample with mean Einstein radius $\overline{\theta_E}= 1\farcs0$.
The overall percentage of modelable lenses increase as the parent-sample mean Einstein radius is increased, from $\sim43\%$ when $\overline{\theta_E}=0\farcs5$ to $\sim99\%$ when $\overline{\theta_E}=2\farcs0$.  This is not surprising because the samples with greater mean Einstein radii have fewer lenses that fail to pass the modelable constraints.

The distribution in mass-density power-law index of the detected and modelable samples reported in Table~\ref{table} were calculated by fitting a Gaussian function to the binned histograms shown in Figure~\ref{histogram}.  The bias in the lens power-law index relative to the parent sample is small, shifted towards slightly shallower lensing potentials in all simulation runs.  This slight bias is introduced by the detection process; the modelable constraints introduced no additional bias.  Furthermore, as the mean Einstein radius of the parent sample was increased, there is a slight but consistent trend in the detected and modelable $\gamma$ bias toward shallower potentials.  This relation is further discussed in \sref{correlation}.
The SLACS-like lens sample bias in $\gamma$ is small enough (0.01 --- 0.03) that it is probably negligible compared to other sources of uncertainty.

\begin{figure*}[t]
\includegraphics[scale=.925]{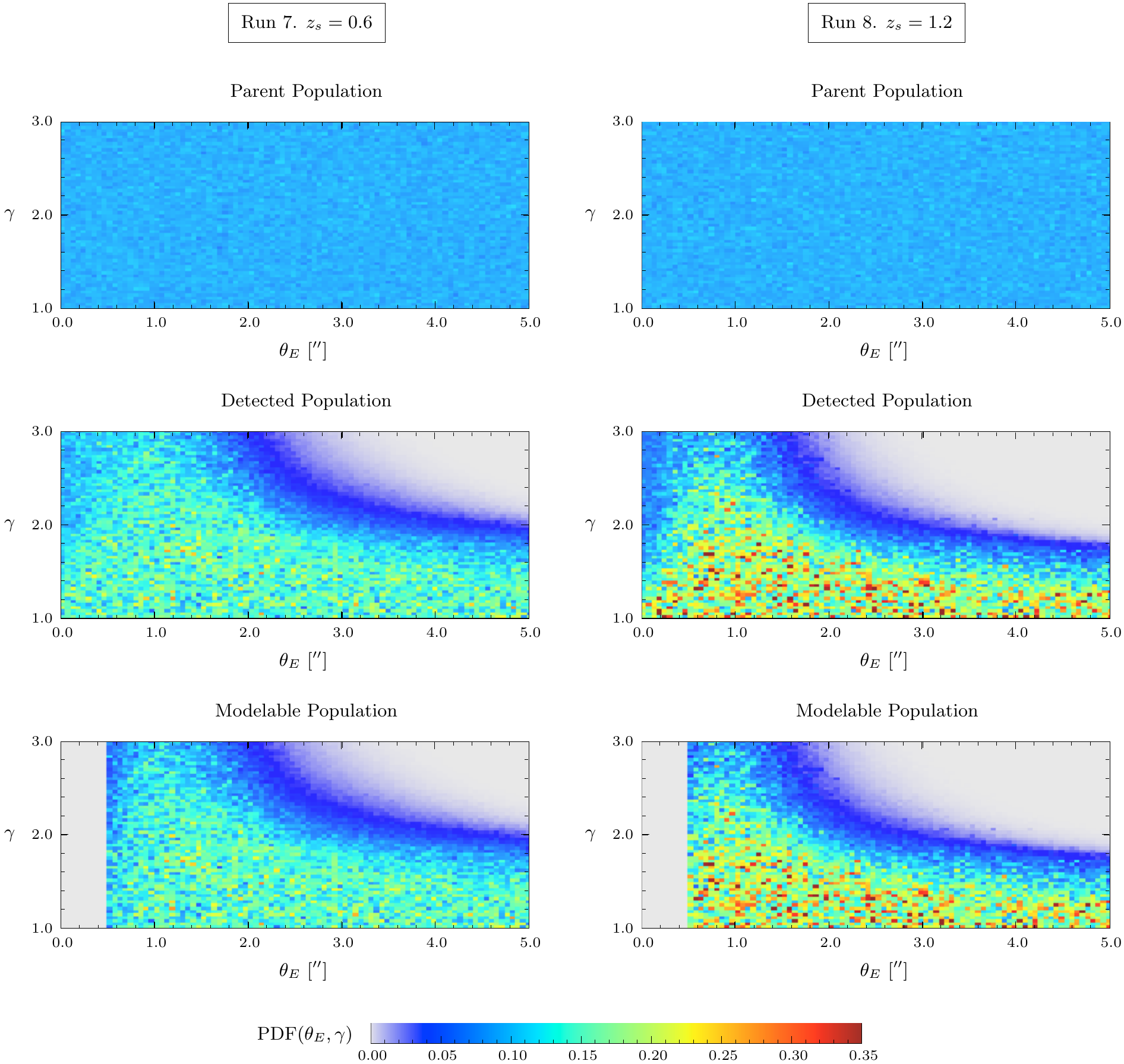}
\figcaption{\label{parcor}Two-dimensional histogram of the correlation between the lens mass-density power-law index, $\gamma$, and lens Einstein radius, $\theta_E$.  The left column corresponds to the mock SLACS simulation (Run 7), with $z_s=0.6$, and the right column corresponds to the mock BELLS simulation (Run 8), with $z_s=1.2$. The top row  shows the parent samples of \(2 \times 10^6\) lenses uniformly distributed across the shown ranges.  The middle row shows the distribution of the detected lenses, and the bottom row shows the modelable lens sample distribution.   The parent and detected histograms are normalized to unit area, whereas the modelable histogram is normalized to the fraction of modelable lenses relative to detected lenses. Note the decrease in detected and modelable lenses with high mass-density power-law indices at large Einstein radii and the cutoff in the modelable lens sample at low Einstein radii due to the modelable criteria.}
\end{figure*}

Finally, we note that for every run of the simulation there is no obvious bias in the lens mass axis ratio, either in terms of detectability or modelability.  A similar result was found by \citet{dobler08}.

\subsection{Dependence on Source Parameters}
Although we don't quantify the detection and modelable biases for the source parameters, it is of interest to understand them qualitatively.  As seen in Figure~\ref{histogram} there is a distinct bias towards smaller source sizes in both the detected and modelable samples in all of the SLACS simulations tested (Runs 1--3).  This result probably arises because higher magnification smaller source sizes can acquire greater [O\,{\sc ii}] line flux within the spectroscopic fiber, which makes them more likely to be detected. For every simulation tested, there is no detection or modelability bias in the source position.

\subsection{Simulation Images}

Figure~\ref{sim_ims} depicts example lens images from the mock SLACS simulations.  It also illustrates the effects of seeing on the detection of lens systems.  As seen in Figure~\ref{sim_ims} our simulation includes false positive lenses, or lenses which are detected but not modelable because they lack a counter image.  Of course our simulation also includes the opposite, lenses which are not detected but are modelable.  These lenses fail to be detected either due to seeing effects, low source luminosity, or the lens images falling outside of the fiber.  It is possible, however, for a system with a large Einstein radius and low mass-density power-law index to be detected and subsequently classified as modelable when just the central, de-magnified image is inside the fiber.  This is seen particularly at small source redshifts and high source luminosities.  This effect makes the relationship between $\theta_E$ and $\gamma$ an interesting one to investigate which we pursue in \sref{correlation}.  Figure~\ref{sim_ims} also includes examples from the simulation which are detected and modelable as well as not detected and not modelable.

\subsection{Mock BELLS Simulation Results}

The results of Runs 4--6 of the simulation with parameters typical of BELLS are displayed in the last three rows of Table~\ref{table} and Figure~\ref{histogram}.  One of the main differences between BELLS and SLACS is the size of the fiber, which we have already seen is an important selection effect and source of bias in spectroscopic strong lens surveys.  This is manifested primarily in a more pronounced bias towards smaller detected Einstein radii and slightly shallower detected mass-density
profiles for the simulation with the largest mean Einstein radius of $\overline{\theta_E}=2\farcs0$.
For the parent samples with smaller mean Einstein radii, the biases in $\gamma$ remain small for
both detected and modelable subsets, and the bias in $\overline{\theta_E}$ is small for
the detected subsample.  The modelable bias in $\overline{\theta_E}$ is zero for the $\overline{\theta_E}=1\farcs0$ BELLS simulation, and is identical to bias in the SLACS simulation
for the $\overline{\theta_E}=0\farcs5$ parent sample.
Another result of the BELLS simulations is that the percentage of detected lenses is lower than that detected in any of the SLACS simulations.  However, the average percentage of modelable lenses is slightly higher than in the SLACS simulations.  Compared to the SLACS simulations, the BELLS simulations also show a much more distinct bias in detecting smaller source sizes (see Figure~\ref{histogram}).  Again as with the SLACS-like simulations, there is no bias in the lens mass axis ratio, detected or modelable.

\subsection{Parameter Correlation}
\label{correlation}

The results of our uniform exploration of the plane of $\theta_E$ and $\gamma$ are presented in Figure~\ref{parcor}.  For both the detected and modelable lenses, as the Einstein radius increases the minimum detectable power-law index decreases monotonically.  This effect is at least partially due to selection effects of spectroscopic surveys (i.e., the finite size of the fiber) which eliminate large separation lenses.  In spite of this, the result implies that large radius, shallow lensing potentials could in principle be detected and modeled by spectroscopic surveys, but that such surveys are blind to large radius, steep lensing potential galaxies.  However, real systems with very large Einstein radii will tend to be of the group or cluster scale (exceeding 400 km\,s$^{-1}$), and will have more complicated lensing potentials: it is unclear whether our simple
power-law model is applicable to such systems.

\section{Conclusions}
\label{conclusion}
We have presented a simple Monte Carlo simulation used to explore the selection biases of spectroscopic strong galaxy--galaxy gravitational lens surveys.  As \citet{dobler08} found and we confirm in this work, the size of the spectroscopic fiber is the most significant selection effect in strong galaxy surveys.  \textit{Detection} biases become important for parent populations with a mean Einstein radius of approximately $2\farcs0$, shifting the detected samples toward relatively smaller Einstein radii and shallower lensing potentials.  \textit{Modelability} biases become important for parent populations with smaller mean Einstein radii (of approximately $0\farcs5$), pushing the modelable subsamples to larger mean Einstein radii but not biasing the average mass-density profile significantly.  We find that a majority of detected lenses are classified as modelable in all the simulations tested, thus explaining the efficiency of follow-up high-resolution imaging of spectroscopically discovered lens candidates.  We find no significant bias in the lens mass axis ratio relative to the
parent population.

An important implication of this work is that, for galaxy-scale
spectroscopically selected lenses,
the bias in the detected and modelable mass-density profile slope relative
to the parent population is relatively small, and hence is not likely to
be a major factor in the interpretation of results
on the mass-density profile of lensing galaxies
such as those of \citep{koopmans06,koopmans09} and \citet{bolton12}.
However, the constraint of lens modelability for populations of lower
mean Einstein radius indicates that the lenses selected from these
populations are biased systematically towards higher velocity dispersions.
One mitigating strategy to correct for this bias would be to include
constraints from \textit{non-lenses} along with constraints from
lenses in the analysis of survey samples, since non-lenses with
identifiable single-source images place upper limits on the velocity
dispersions of their foreground galaxies.

The comparison between SLACS-like and BELLS-like simulations shows that the detailed
parameters of the spectroscopic survey are important to determining the relevant
details of the spectroscopic lens selection function.
Nevertheless, there is significant commonality in the trends of selection
bias between these two survey configurations.
In view of the substantially uniform detection
sensitivity of both SLACS-like and BELLS-like
surveys across a broad range of Einstein radii and mass-density
profile slopes, such surveys can be used to select representative
samples of field galaxies for study through gravitational lensing,
with biases that can be quantified and corrected.  At group and cluster
lens-mass scales, spectroscopic selection becomes much less efficient
and also much more sensitive to the parameters of the mass-density profile.

The bias of spectroscopically selected lenses toward sources of smaller physical size,
seen for both SLACS and BELLS simulations (with much greater magnitude in the latter)
has important implications for the study of the lensed galaxy population discovered
in these surveys.  This effect can explain the mass--size offset
seen in the work of \citet{newton11} for lensed emission-line galaxies
relative to broadband-selected un-lensed galaxies at comparable magnitudes and redshifts.
This bias of course affords the advantage of studying a systematically smaller
and fainter population of galaxies than would be accessible without the
strong lensing effect.

This work has highlighted the salient features of the selection
function of spectroscopic gravitational lens surveys.
Quantitative correction for the biases identified by this work
will require simulations that account for varying source-galaxy redshifts
and for multiple parent Einstein radius (i.e., velocity-dispersion)
distributions.  In future work we will integrate detailed implementations
of our Monte Carlo simulations into the analysis of population distribution
parameters for the combined SLACS and BELLS gravitational lens data set,
thus allowing a more robust generalization from the physical
nature of spectroscopic lens samples to the population of early-type
field galaxies as a whole.

\acknowledgments 

Funding for the SDSS and SDSS-II has been provided by the Alfred P. Sloan Foundation, the Participating Institutions, the National Science Foundation, the U.S. Department of Energy, the National Aeronautics and Space Administration, the Japanese Monbukagakusho, the Max Planck Society, and the Higher Education Funding Council for England. The SDSS Web Site is \url{http://www.sdss.org/}.

The SDSS is managed by the Astrophysical Research Consortium for the Participating Institutions. The Participating Institutions are the American Museum of Natural History, Astrophysical Institute Potsdam, University of Basel, University of Cambridge, Case Western Reserve University, University of Chicago, Drexel University, Fermilab, the Institute for Advanced Study, the Japan Participation Group, Johns Hopkins University, the Joint Institute for Nuclear Astrophysics, the Kavli Institute for Particle Astrophysics and Cosmology, the Korean Scientist Group, the Chinese Academy of Sciences (LAMOST), Los Alamos National Laboratory, the Max-Planck-Institute for Astronomy (MPIA), the Max-Planck-Institute for Astrophysics (MPA), New Mexico State University, Ohio State University, University of Pittsburgh, University of Portsmouth, Princeton University, the United States Naval Observatory, and the University of Washington.

\bibliography{paperNotes}

\end{document}